\documentclass{osa-article}
\journal{oe}

\usepackage{cite}
\usepackage{mathrsfs}
\usepackage{amsmath}
\usepackage{array}
\usepackage[caption=false,font=footnotesize]{subfig}
\usepackage{color,soul}

\begin{document}

\title{Demonstrating Delay-based Reservoir Computing Using a Compact Photonic Integrated Chip}

\author{Krishan~Harkhoe, \authormark{1,*} Guy~Verschaffelt, \authormark{1} Andrew~Katumba, \authormark{2} Peter~Bienstman, \authormark{2} and Guy~Van der Sande  \authormark{1}}
\address{\authormark{1}Applied Physics Research Group, Vrije Universiteit Brussel, Pleinlaan 2, 1050 Brussels, Belgium \\
\authormark{2}Photonics Research Group, Department of Information Technology, Ghent University-IMEC, Technologiepark Zwijnaarde 126, 9052 Ghent, Belgium}

\email{\authormark{*}kharkhoe@vub.be} 


\vskip 1em
\textbf{Abstract: ~}Photonic delay-based reservoir computing (RC) has gained considerable attention lately, as it allows for simple technological implementations of the RC concept that can operate at high speed. In this paper, we discuss a practical, compact and robust implementation of photonic delay-based RC, by integrating a laser and a 5.4~cm delay line on an InP photonic integrated circuit. We demonstrate the operation of this chip with 23 nodes at a speed of 0.87~GSa/s, showing performances that is similar to previous non-integrated delay-based setups. We also investigate two other post-processing methods to obtain more nodes in the output layer. We show that these methods improve the performance drastically, without compromising the computation speed.

\section{Introduction}\label{sec:Intro}
The concept of reservoir computing (RC), a paradigm within neuromorphic computing, offers a framework to exploit the transient dynamics within a recurrent neural network for performing useful computation. It has been demonstrated to have state-of-the-art performance for a range of tasks that are notoriously hard to solve by algorithmic approaches, e.g., speech and pattern recognition and nonlinear control. RC simplifies the training procedure for recurrent neural networks, by keeping the neural network fixed and relying on a trained output layer that consists of a linear combination of network states to generate the desired output signals. Hence, during training only the connections from the network to the output layer are trained. The fixed network is called the reservoir and can actually be any dynamical system with a high dimensional state space. Due to this simplification, RC rekindled neuromorphic computing activities in photonics. Today, multiple photonic RC systems can provide a practical yet powerful hardware substrate for neuromorphic computing \cite{van2017advances}. Some examples include a network of semiconductor optical amplifiers \cite{vandoorne2008toward,vandoorne2011parallel}, an integrated passive silicon circuit forming a very complex and random interferometer, with nonlinearity introduced in the readout stage \cite{vandoorne2014experimental} and a semiconductor laser network based on diffractive coupling \cite{brunner2015reconfigurable}.

The concept of delay-based RC, using only a single nonlinear node with delayed feedback, was introduced some years ago by Appeltant \textit{et al.} \cite{appeltant2011information} as a means of minimizing the expected hardware complexity in photonic systems. The first working prototype was developed in electronics in 2011 by Appeltant \textit{et al.} \cite{appeltant2011information} and several performant optical systems followed quickly after that \cite{paquot2010reservoir,paquot2012optoelectronic}, one of which is based on a semiconductor laser with external optical feedback \cite{brunner2013parallel}.

Delay-based RC offers a simple technological route to implement photonic neuromorphic computation. Its operation boils down to a time-multiplexing with the delay arising from propagation in the external feedback loop, limiting the resulting processing speed. As most optical setups end up to be bulky employing long fiber loops or free-space optics, the processing speeds are limited in the range of kSa/s to tens of MSa/s \cite{paquot2012optoelectronic,brunner2013parallel}. To increase the processing speed of delay-based reservoir computing using a semiconductor laser with delayed optical feedback, one can integrate the laser and the delay both on the same photonic chip. In this way, by using a waveguide structure with a compact footprint, an external cavity structure can be implemented which is small enough to reach high processing speeds, yet still long enough to have sufficient dimensionality for good computing performance. In the long term, this integrated approach will lead to a robust and low-cost design. 

Recently, Takano \textit{et al.} \cite{takano2018compact} have presented a photonic integrated circuit (PIC) consisting of a distributed-feedback semiconductor laser, a semiconductor optical amplifier (SOA), a phase modulator, a short passive waveguide, and an external mirror for optical feedback. The external cavity length in this system reached 10.6mm, corresponding to a round-trip delay time of $254$~ps. However, only six virtual nodes could be stored within the delay line with node-spacings of $40$~ps, not enough for good computational performance. This necessitated the authors to use masks with duration of multiple delay times, which slows down the computation speed. 

Our goal is to show that a delay-based reservoir computer can be built using an indium-phosphide PIC, that combines active and passive elements and is built on the JePPIX platform \cite{leijtens2011jeppix}. The PIC integrates a semiconductor laser with an external cavity of $5.4$~cm, which corresponds to a round trip time of $1170$~ps. This allows for $23$ nodes and a processing speed of $0.87$~GSa/s. The longer waveguide based external cavity will also have more loss associated to it. Therefore, we will address in this work the question if amplification in the external cavity is needed or not. Contrary to other works \cite{brunner2013parallel, bueno2017conditions, takano2018compact}, the semiconductor laser itself will be driven far above solitary lasing threshold to benefit from a better signal to noise ratio in the read-out, as well as faster internal dynamics. Finally, we will introduce post-processing schemes that do not penalize computational speed.

In Section II, we describe the experimental setup as well as the pre- and post-processing of data. In section III we present and discuss the results for the different post-processing schemes. We also discuss the linear and nonlinear memory capacity of the system in section III.

\section{Experimental setup}\label{sec:Exp_setup}

\begin{figure}[!t]
\centering
\includegraphics[width=.7\columnwidth]{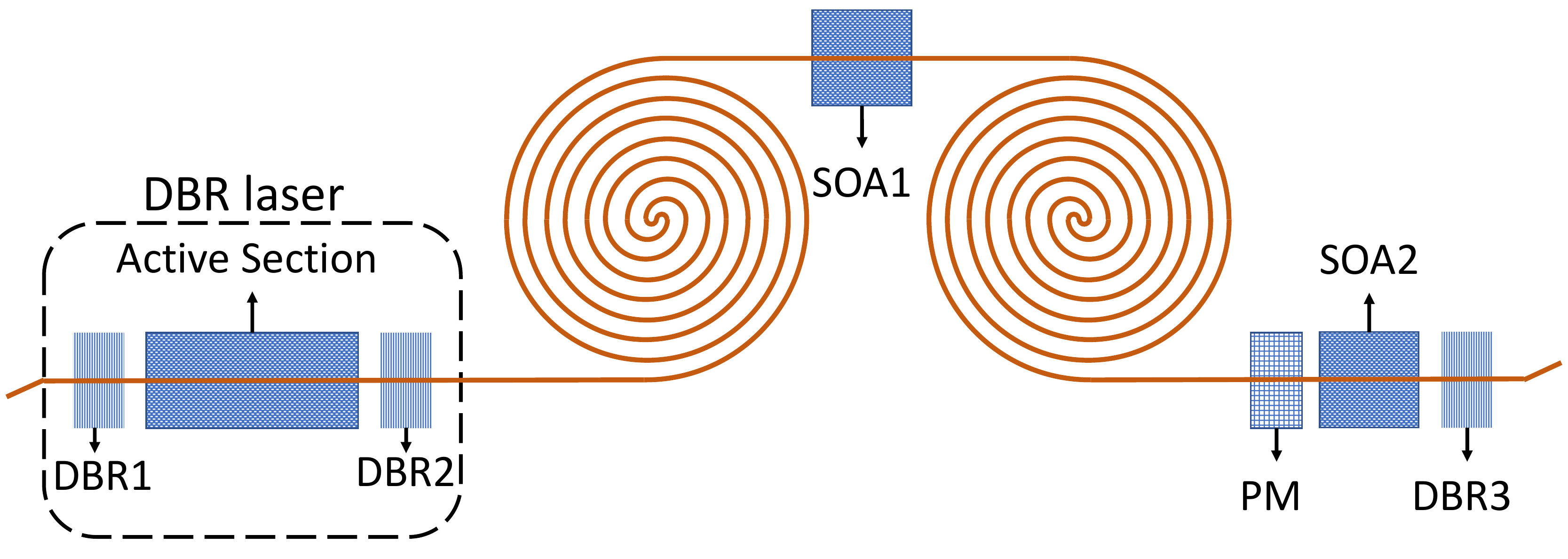}
\caption{Schematic depiction of our InP-based photonic integrated circuit (PIC). The PIC consists of a laser structure followed by a delay line of $5.4$~cm. DBR: Distributed Bragg Reflector, SOA: Semiconductor Optical Amplifier, PM: Phase Modulator. }\label{fig:Chip}
\end{figure}

A schematic of our integrated device is shown in Fig.~\ref{fig:Chip}. It consists of a distributed Bragg reflector (DBR) laser structure and two spiral waveguides comprising the delay line. Two semiconductor optical amplifiers (SOA) are placed along the delay line to tune the feedback strength. A phase modulator is available to tune the feedback phase. At the end of the delay line a DBR element completes the feedback loop by reflection. This on-chip feedback loop has a round-trip time of $\tau = 1170$~ps.

The device covers the whole $6$mm width of the chip and has one optical input/output port on each side. The ports are angled with respect to the chip edge to minimize reflection. We employed lensed fibers to send optical signals in/out of these ports and a total of five electrical DC probes to operate the device. The first probe ($I_{DBR1}$) was placed on the left DBR of the laser structure, in order to tune the spectral output of the laser. The second probe ($I_{L}$) acted to supply the pump current to the laser. The following two probes ($I_{SOA1}$, $I_{SOA2}$) supplied current to the SOAs along the feedback line and the last probe ($I_{DBR3}$) tuned the reflection spectrum of the DBR at the end of the feedback line. The active and SOA sections could be pumped up to a current of $40$~mA, whereas the tuning currents of the DBRs could only be driven up to $10$~mA.

\begin{figure}[!t]
\centering
\includegraphics[width=.7\columnwidth]{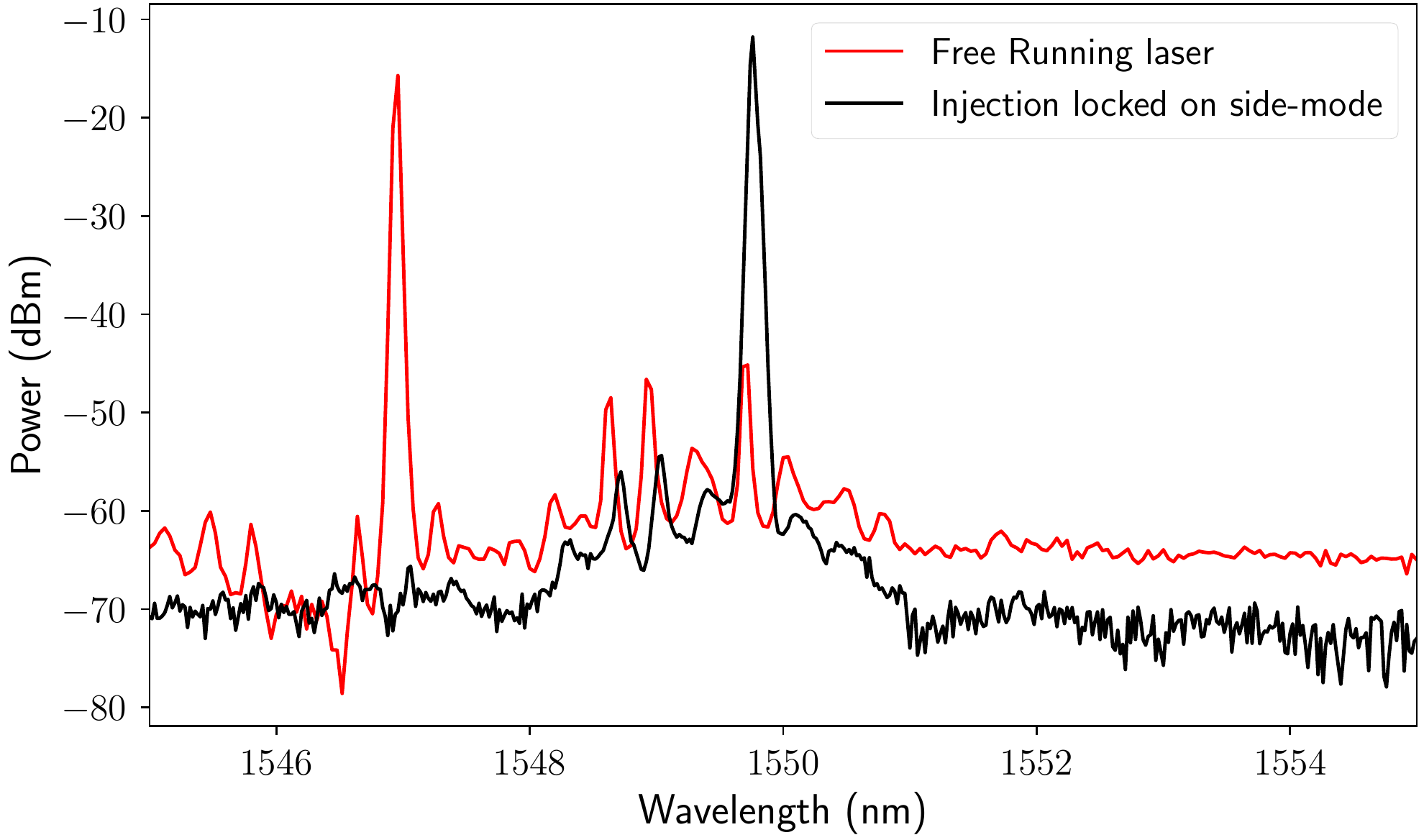}
\caption{The optical spectrum of the free running laser (red) superimposed on the spectrum of the injection locked laser (black). Injection locking was achieved at one of the side-modes at $1549.60$~nm. }\label{fig:Spectrum}
\end{figure}

The DBR laser has a threshold current of $15$~mA. The spectrum of the free running laser is shown in red in Fig.~\ref{fig:Spectrum}, when pumped at $40$~mA and measured at the left output waveguide in Fig.~\ref{fig:Chip}. The free running lasing wavelength is centered at $1546.91$~nm. In our setup, the on-chip laser can lock on the injection at the free running lasing wavelength or one of the side-modes, depending on the injected wavelength. It turned out that the RC performance is best when the injected field's wavelength is close to a side mode, as shown by the black spectrum in Fig.~\ref{fig:Spectrum}. Targeting the side mode allows for a higher injected power as the reflection of DBR1 is lower at the wavelength of the side-mode. Furthermore, DBR2 has a higher transmission for side modes than for the free running lasing wavelength. Injection locking on the side-mode in Fig.~\ref{fig:Spectrum} is achieved at a wavelength of $1549.60$~nm and the following DC probes configuration: $I_{DBR1} = 8.28$~mA, $I_{DBR3} = 1$~mA, and $I_L = I_{SOA1} = I_{SOA2} = 40$~mA. The on-chip spectral parameters are not changed hereafter, meaning that the current supply to the two DBRs is not changed throughout the paper.

\begin{figure}[!t]
\centering
\includegraphics[width=.7\columnwidth]{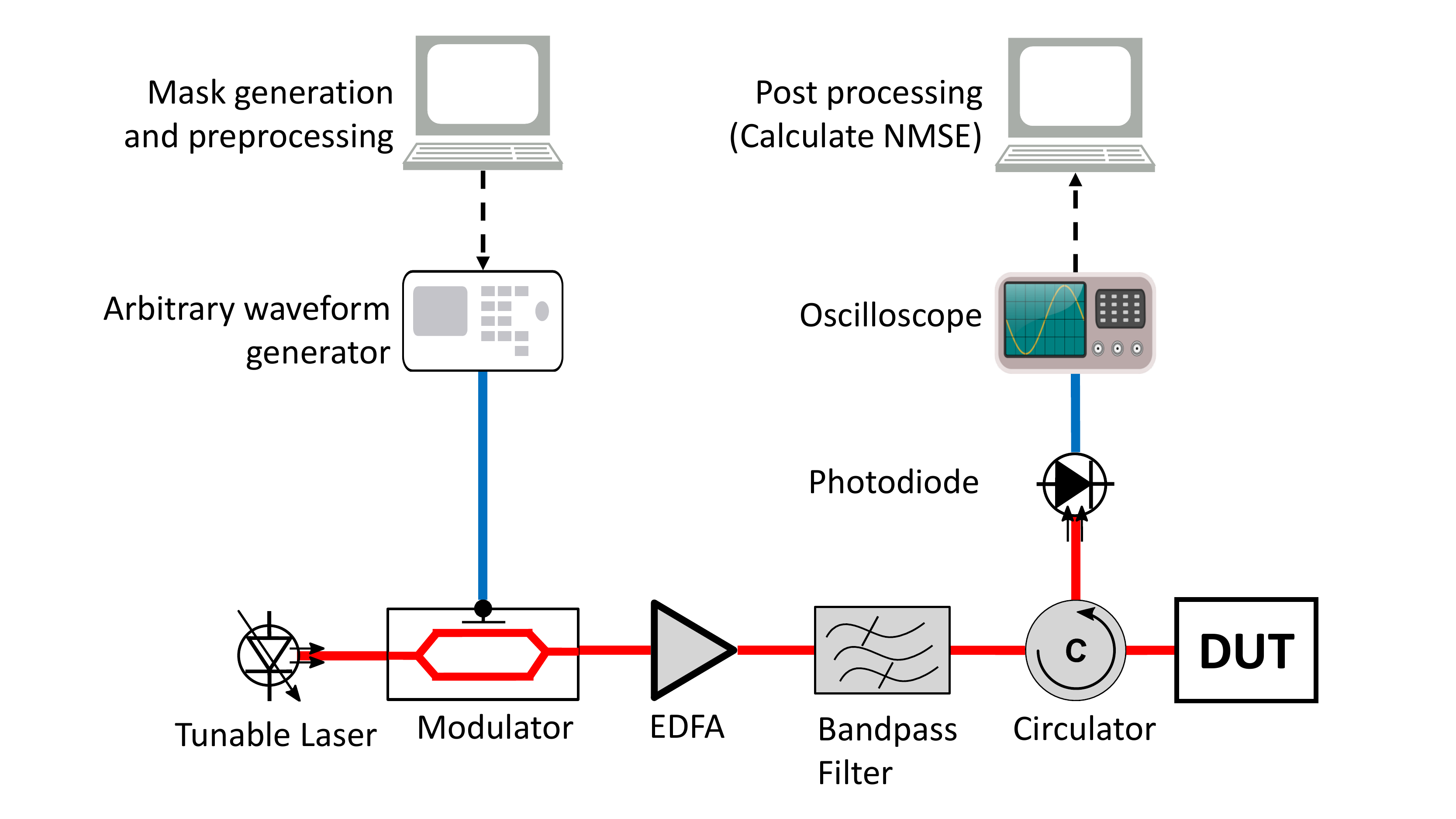}
\caption{Schematic depiction of the setup used to measure the performance of our integrated delay-based reservoir computer. The device under test (DUT) is the PIC shown in Fig.~\ref{fig:Chip}}\label{fig:Exp_setup}
\end{figure}

To test the RC performance of the laser integrated with a feedback loop, the setup shown in Fig.~\ref{fig:Exp_setup} is used. We use a wavelength tunable CW laser to create the optical injection signal. The wavelength of this laser is set close to $1549.6$~nm, but we still allow for a small detuning between the injection wavelength and the wavelength of the targeted side-mode of the laser. The CW light beam of the tunable laser is modulated using a $40$GHz Mach-Zehnder modulator (iXblue MX-LN-40). This modulator is driven electrically by a $25$GHz Arbitrary Waveform Generator (Keysight M8195A) set at a sample speed of $60$~GSa/s.

We employ the time-multiplexing scheme, as introduced in \cite{appeltant2011information}, where the duration of one data sample matches the $1170$~ps delay time. Note that there have been numerical and experimental studies, where the duration of a data sample does not match the delay time \cite{paquot2012optoelectronic, duport2016fully}. We, however, do not target this working regime. 

Any input data sample $u_i$, in our case originating from a discrete timeseries, is held constant for the duration of one delay time $\tau$. We then multiply this piecewise constant stream $U(t)$ with a piecewise constant mask $M(t)$ (that is periodic with a period of $\tau$) to obtain the masked input stream $J(t)$. The piecewise constant levels of stream $J(t)$ define the position of the virtual nodes equally spread over the delay line. It has been shown numerically \cite{nguimdo2014fast} that the node separation, when using a semiconductor laser with delayed feedback, can be as short as a few tens of ps. As the sample rate of the AWG is set to $60$~GSa/s, we use three AWG samples to define one mask node, leading to a mask node separation of $\theta_M = 50$~ps such that $23$ nodes fit within one round-trip in the delay loop. We thus generate a random mask with $N_M = 23$ mask nodes with three possible values $[0,0.5,1]$. 
In our case the length of the mask is $20$~ps shorter than the delay time, which is hard to match in practice. We believe this desynchronization will not adversely affect the performance of the RC scheme, since the mismatch is smaller than the node separation and we can accurately split the reservoir output in the readout layer.

\begin{figure*}[!t]
\centering
\includegraphics[width=.99\textwidth]{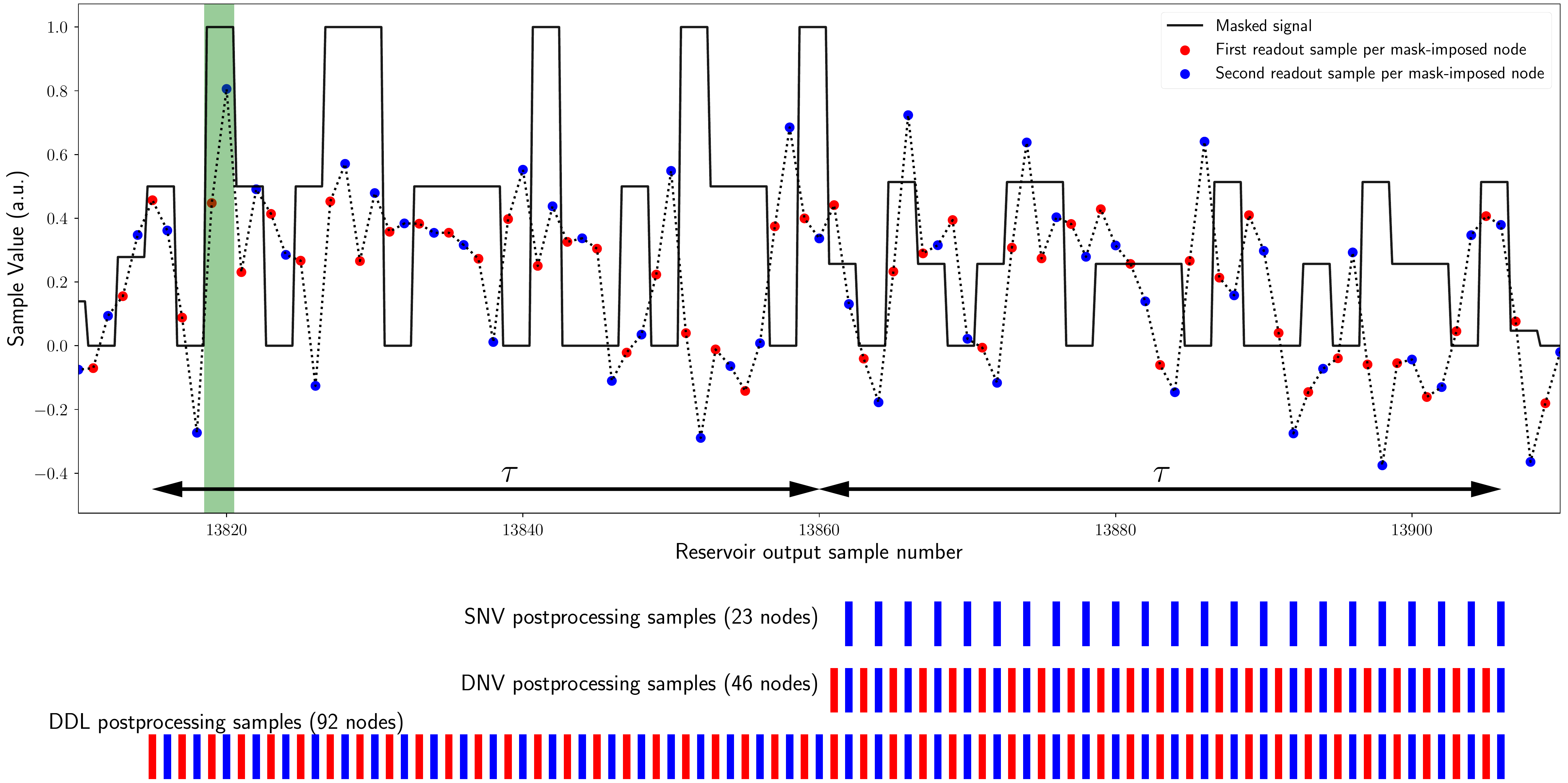}
\caption{This figure illustrates the different timescales in the experimental setup. A masked data signal (black full line) is superimposed on the reservoir output (black dotted line). The green shaded region corresponds to the duration of one mask-imposed node and each mask node has two readout samples. The first and second readout samples per mask-imposed node are shown in red and blue respectively. 
The length of one delay time $\tau$ is indicated by a black arrow. Beneath the time traces, we used colored tickmarks to indicate the samples that are taken into account for the three post-processing routines.}\label{fig:Traces}
\end{figure*}

The modulated optical signal is next amplified in Fig.~\ref{fig:Exp_setup} using an Erbium doped fiber amplifier (Keopsys CEFA-C-BO-HP-B203). The broadband spontaneous emission noise, added to the optical signal by the amplifier, is removed by sending the light beam through an optical bandpass filter, that is centered around the injection signal's wavelength. The filtered signal is then fed into the laser using a circulator connected to a lensed fiber. The response of the laser is collected at the third port of the circulator and measured using an opto-electronic detector connected to a $63$GHz real-time oscilloscope. The sampling rate of the oscilloscope was set to $40$~GSa/s. This means that each mask node, with a duration of $50$~ps, has 2 corresponding samples in the read-out signal. This is illustrated in Fig.~\ref{fig:Traces}, where we show an overlay of the masked signal and the reservoir output. The green shaded area corresponds to one mask node ($50$~ps) and we see two read-out samples in this shaded region. 

\subsection*{Benchmarking and performance indicator}
The benchmark task we have used, is the one-step-ahead forecast of a laser-generated dataset from the Santa Fe timeseries prediction competition \cite{StanfordUniversity}. The set consists of $9092$ points, of which we used only the first $5000$. From these $5000$ points, the first $3500$ points were used for training. The last $1500$ points were allocated for testing the performance on unseen data. We used the normalized mean square error ($NMSE$) as performance indicator, which is defined as: 
\begin{equation}
NMSE(y,y_{exp}) = \frac{\left\langle    || y(n) -y_{exp}(n)||^2   \right\rangle }{\left\langle    || y_{exp}(n) -  \left\langle y_{exp}(n) \right\rangle   ||^2   \right\rangle}, 
\end{equation}
where $y$ is the predicted value and $y_{exp}$ is the expected value, $n$ is a discrete time index and the symbols $||...||$ and $ \left\langle ... \right\rangle$ stand for the norm and the average respectively. The $NMSE$ is always a positive value, with lower $NMSE$ values corresponding to better performances.

\subsection*{Post-processing}
Photonic systems are inherently noisy systems, which usually is helpful to avoid overfitting the reservoir to training data. However, in our experiment we were limited by a very noisy photodiode, which lead to a relatively low SNR of 6~dB. To increase the SNR in the output layer, we have recorded several sequential repetitions of the same input signal and averaged them, as was also done in Ref.\ \cite{brunner2013parallel}. We have analyzed these  repetitions and found the response of the laser each time to be consistent within the noise level. With the SNR now improved to 21~dB, we performed the training and testing on the average of these traces.

We performed three different post-processing routines. Recall that we obtain two output samples per mask-imposed node in the read-out layer. In the first post-processing routine, we only take the last sample per mask node. This means that the virtual node distance $\theta_V$ equals the node distance $\theta_M$ as imposed by the mask. This is the conventional post-processing routine in delay-based reservoir computing. 

The second routine utilizes both samples of each mask node and treats them as separate nodes, such that the number of virtual nodes is twice the number of nodes imposed by the mask, $N_V=2N_M$ and $\theta_V=\theta_M/2$. Note that this second routine is also used by Takano \textit{et al.}\cite{takano2018compact}. Fig.~\ref{fig:Traces} shows that the two samples per mask node do not necessarily have the same value due to the transient response of the laser. That is why we presume that the second post-processing routine might have a richer state space to function as a reservoir computer, than the single node value post-processing routine. 

In the last routine, we take the reservoir states over a duration of $2\tau$ and use all detector samples per mask-imposed node. The output layer in this case consists of virtual nodes from the last two masked input values, in contrast to the other two routines, where the virtual nodes from the last masked input value is being considered. In this case we get a virtual node separation $\theta_V=\theta_M/2$, since both output samples per mask-imposed node are taken to form the output layer. Furthermore, we get four times more virtual nodes than mask-imposed nodes, $N_V=4N_M$. Note that we do not change anything in the preprocessing (masking), so our computation speed remains the same.

We will refer to the three routines as single node value (SNV) post-processing, double node values (DNV) post-processing and double readout length (DRL) post-processing, respectively. The nodes taken into account for each post-processing routine are illustrated in Fig.~\ref{fig:Traces}, together with a readout timetrace. 

\section{Results}\label{sec:Results}
\subsection*{SNV post-processing}\label{subsec:SNV}

\begin{figure*}[t]
\centering
\includegraphics[width=.99\textwidth]{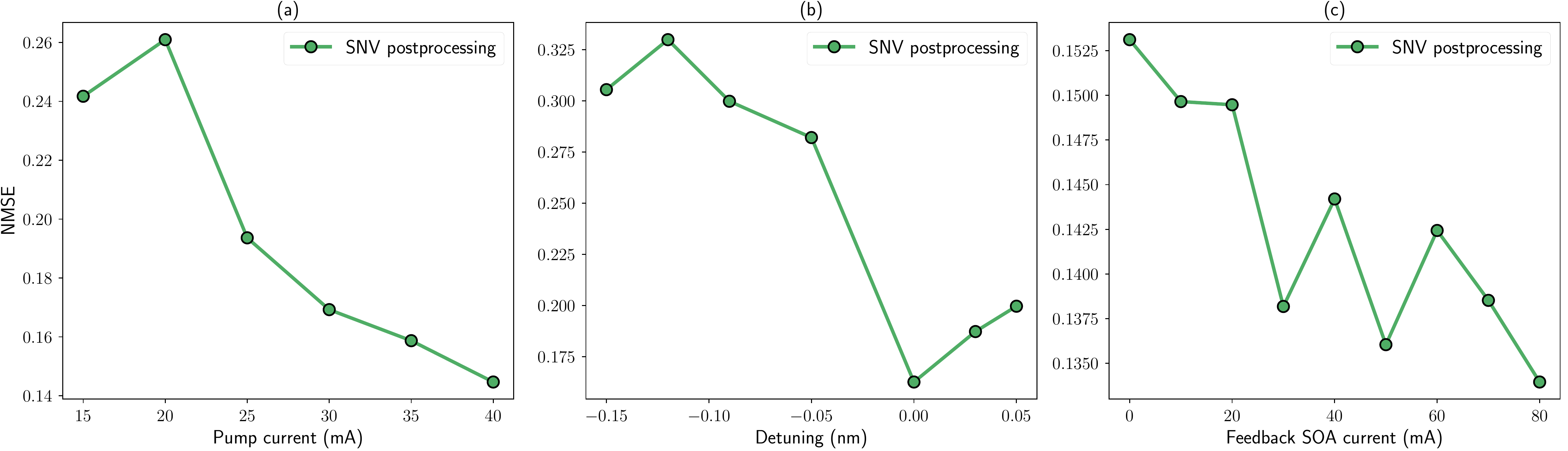}
\caption{The performance of the RC scheme as (a) the laser pump current is increased, (b) the wavelength detuning of the injected beam is varied and (c) as the sum of the current to SOA1 and SOA2 is increased. The performance is expressed by the normalized mean square error ($NMSE$). The results obtained here are from the single node value (SNV) post-processing scheme.}\label{fig:SNV}
\end{figure*}

We will first discuss the results obtained from the single node value (SNV) post-processing routine. The first parameter that we scanned in the experiments, was the pump current of the laser and the result is shown in Fig.~\ref{fig:SNV}(a). The general trend we can observe here, is that the reservoir performs better at higher pump currents. Other studies, such as Bueno \textit{et al.}\cite{bueno2017conditions}, Nguimdo \textit{et al.} \cite{nguimdo2014fast} and Takano \textit{et al.}\cite{takano2018compact}, have always operated the laser in regions close to the solitary laser threshold and found that the performance worsens as the pump current increases. However, they typically scan the pump current over $0.9-1.1 I_{threshold}$, whereas we investigate in the range of $1.0-2.5_{threshold}$. These previous studies achieve $NMSE$ values around $0.1$ for the same Santa Fe timeseries prediction task. At threshold (pump current= $15$~mA), we achieve an $NSME=0.24$. As observed by the aforementioned studies, we see a slight increase at $20$~mA in the $NMSE$, but as we increase the pump current even further we see that the $NMSE$ drops towards $0.14$ at maximum pump current. We believe that by locking on a side-mode, we are able to get more injected power through the DBR, which in its turn stabilizes the laser at higher pump currents. One advantage of pumping the reservoir at currents well above the laser's threshold is a better signal to noise ratio, which leads to a more consistent read-out layer.

The second parameter that we scanned, was the wavelength detuning between the injected beam and the targeted side-mode. The laser is pumped at $I_L =40$~mA and the two SOAs are also supplied with their maximum current of $40$~mA. The result of the scan can be seen in Fig.~\ref{fig:SNV}(b). At a detuning equal to zero, we observe the lowest $NMSE$. When the magnitude of the detuning increases, we see that the performance worsens. This result is in line with previous experimental study by Bueno \textit{et al.} \cite{bueno2017conditions}, who observed the highest consistency at full locking, but better memory capacity at partial locking. The detuning range in our experiments matches the window of full locking as seen in Ref.~\cite{bueno2017conditions}. Going beyond this detuning range, the laser starts to lock to the next side-mode.

Lastly, we vary the feedback strength by varying the current supplied to the two SOAs along the feedback line. The laser is pumped at $40$~mA and the injection wavelength is set at a detuning of $0$~nm, such that we achieve the optimal setting for those parameters. The result of the feedback scan is shown in Fig.~\ref{fig:SNV}(c), where the sum of the currents supplied to the two SOAs is placed along the x-axis. We see as general trend here that the performance improves as the feedback from the delay line is increased. The rather non-monotonous progress of the measured $NMSE$ values can be attributed to changes in the feedback phase. As the current of the SOAs increases, the path length of the delay line changes due to thermal effects. Due to practical constraints, we did not use a sixth probe to adjust the feedback phase. 

If the improvement of $NMSE$ is compared over the three scans, we see that changing the feedback strength is not as significant as the other two parameters.  Feedback strength is generally, but not exclusively, related to the memory capacity of a delay-based reservoir \cite{bueno2017conditions}. We have estimated that the overall round trip loss in the external cavity (not counting the reflectance of DBR3) amounts to -26~dB. Comparing the power levels when both SOA1 and SOA2 are pumped at 40~mA, implies a reduction of this loss to -6~dB. With a transmittance of 0.8 for DBR2 and a reflectance of DBR3 of about 0.3, we can estimate that the feedback strength ranges from 0.4 (Both SOA1 and SOA2 unpumped) to 40~ns$^{-1}$ (Both SOAs pumped at 40mA each). In the numerical simulations presented in Ref.~\cite{nguimdo2014fast}, feedback strengths as low as 1~ns$^{-1}$ have been shown to lead to good performance on the Santa Fe benchmark task. So, it seems that even without additional amplification in the delay line, we have sufficient feedback strength for a reasonable computation, explaining why our RC-setup works even when the SOAs are not pumped.

\begin{figure*}[!t]
\centering
\includegraphics[width=.99\textwidth]{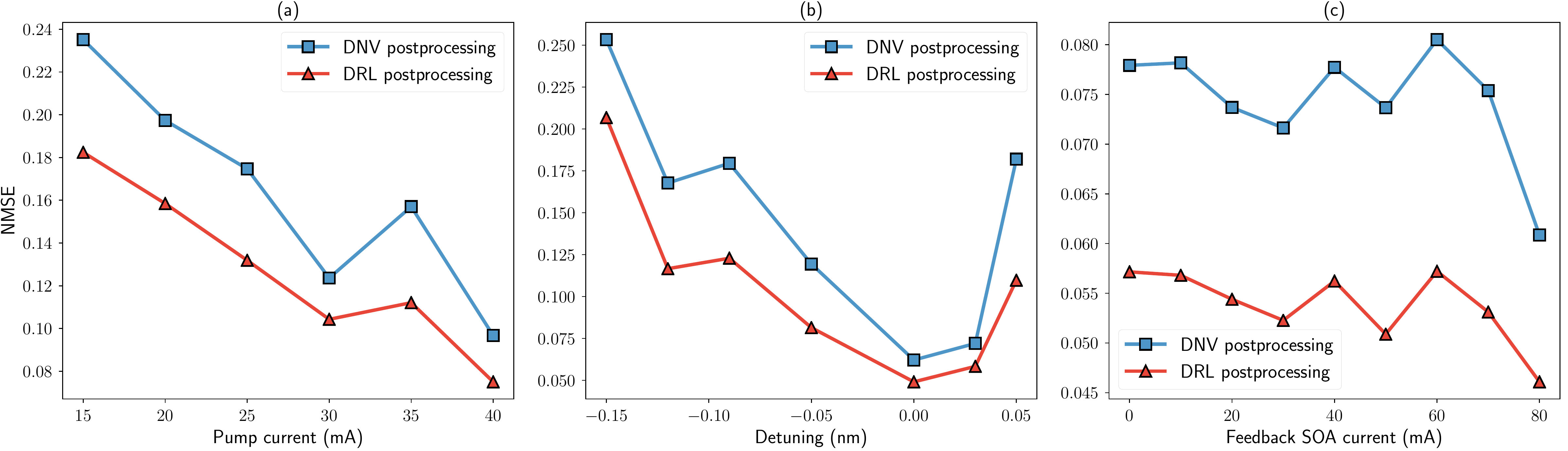}
\caption{The performance of the RC scheme as (a) the laser pump current is increased, (b) the wavelength detuning of the injected beam is varied and (c) as the sum of the current to SOA1 and SOA2 is increased. The results obtained here are from the double node value (DNV) post-processing scheme (where both samples per mask-imposed node are utilized) and the double readout length (DRL) post-processing scheme (where the nodes over the duration of two delay times are considered for processing). The performance is expressed by the normalized mean square error ($NMSE$).}\label{fig:DNV_DRL}
\end{figure*}
\subsection*{DNV and DRL post-processing}\label{subsec:DNV_DRL}
We performed the double node values (DNV) and double readout length (DRL) post-processing routines on the same reservoir output that was used to obtain Fig.~\ref{fig:SNV}. The results are shown in Fig.~\ref{fig:DNV_DRL}.
In general we see the same trends for Fig.~\ref{fig:DNV_DRL} (a) as in Fig.~\ref{fig:SNV} (a), i.e. the performance improves with increasing pump current.
For Fig.~\ref{fig:DNV_DRL} (b) we find the best performance again at zero detuning. The performance degrades with increasing magnitude of detuning. However, the change in performance is less dramatic as compared to Fig.~\ref{fig:SNV}(b). Compared to the SNV routine, the number of virtual nodes in the DNV post-processing is twice as large and in the DRL routine it is four times larger. This larger state space is able to compensate for the consistency that is lost as the injected wavelength moves away from the zero detuning. 
Fig.~\ref{fig:DNV_DRL}(c) illustrates how the performance of the reservoir improves as the feedback is increased. Again we see the non-monotonous progress of the curve, which we believe arises due to additional phase changes along the feedback line as the SOA currents are increased. Again the window of improvement of $NMSE$ due to feedback is less than when the pump-current or the detuning is varied. As discussed earlier for Fig.~\ref{fig:SNV}(c), we believe this is either because the Santa Fe timeseries forecast relies less on the memory capacity or that the feedback from the delay line is already sufficient at zero feedback SOA current. 

A comparison of the $NMSE$ values in Fig.~\ref{fig:SNV} and \ref{fig:DNV_DRL} shows that the performance improves considerably, when we switch from SNV to DNV post-processing routine. The DRL routine consistently outperforms the other two routines on all parameter sweeps. The best $NMSE$ we achieved with the SNV routine is $0.134$, for the DNV routine this drops to $0.062$ and even lowers for the DRL routine to $0.049$. 

These results are in accordance with our expectations. In the SNV routine, we only take one readout sample per mask-imposed node, as is done in most conventional delay-based reservoir computing. The integrated setup we present is doing quite a good job, taking into account that it only consists of $23$ neurons and still giving a best $NMSE$ of $0.135$ at computing speeds of $0.87$~GSa/s. This is in the same range as obtained by Paquot \textit{et al.}\cite{paquot2012optoelectronic} with an optoelectronic setup with $50$ virtual nodes at computing speeds of $0.48$kSa/s. Takano \textit{et al.} obtained a best performance around $NMSE=0.086$, with an integrated setup with $124$ virtual nodes achieving computation speeds of $0.80$~GSa/s. This integrated setup, with a mask length that equals multiple delay times, has a smaller footprint and performs better than our conventional SNV post-processing routine, but requires much more pre- and post-processing in comparison.

When we use our DNV and DRL post-processing routines, the performance improves significantly over all scanned regions. With a best performance of $NMSE=0.062$ for the DNV and $NMSE=0.049$ for the DRL routine, we managed to outperform previous setups. Note that the two latter post-processing routines we used, do not alter the computation speed, as the reservoir keeps running with the same delay line and mask length. It is the mask length that determines the computation speed.

\subsection*{Memory Capacity}\label{subsec:MemCap}
\begin{figure}[!t]
\centering
\includegraphics[width=.99\textwidth]{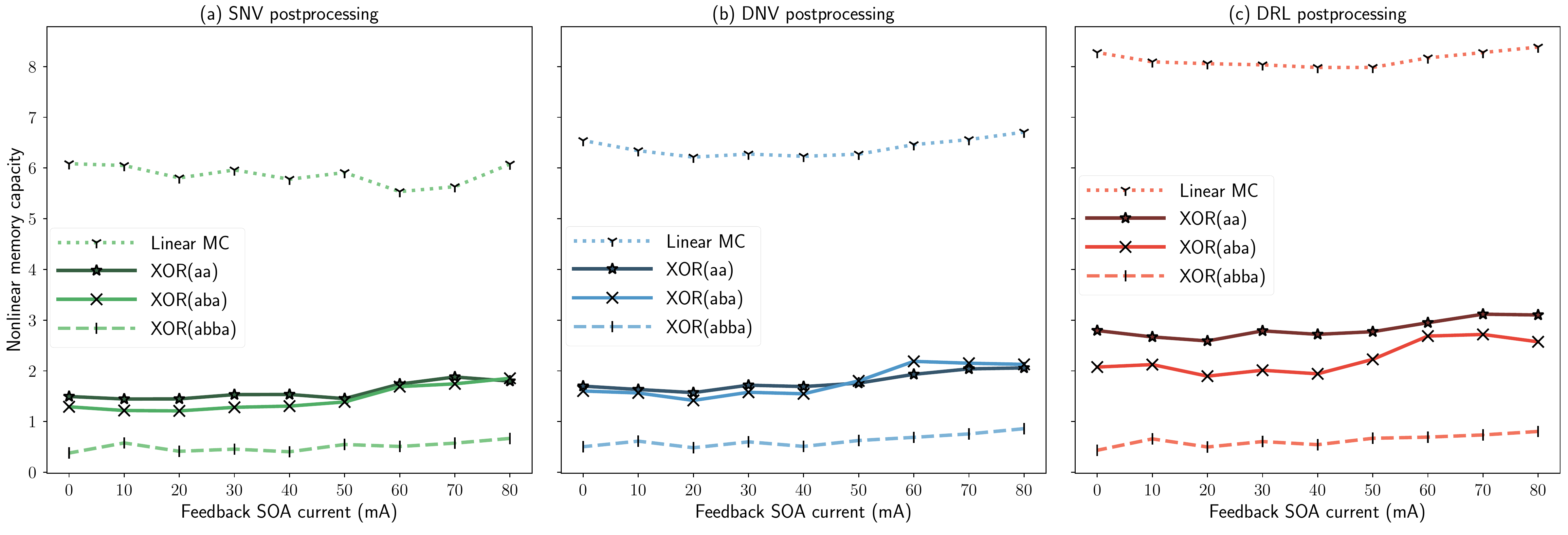}
\caption{The linear and nonlinear memory capacities are shown here for the three post-processing routines ((a) SNV, (b) DNV, (c) DRL) as the feedback strength is increased.}\label{fig:MemCap_Var}
\end{figure}

The results discussed above suggest that the one-step-ahead forecast of the Santa Fe timeseries is not strongly influenced by the memory capacity of the reservoir. Other computational tasks, however, are known to require a substantial amount of memory. Therefore, we also want to test the memory capacity of our integrated system. A measure for linear short-term memory capacity has been introduced in \cite{Jaeger_shortterm} for Echo State Networks. This measure has been employed for reservoir computing schemes, for example in \cite{bueno2017conditions, takano2018compact}. The capacity of a reservoir to recall an input that was fed $i$ samples before, is defined as follows:
\begin{equation}
    m_{i} = \frac{cov^2\left(y_i(n),y_{exp}(n-i)\right)}{\sigma_{y_i}^2 \sigma_{y_{exp}}^2},
\end{equation}
where $y_{exp}(n-i)$ is the input data shifted by $i$ samples, $y_{i}(n)$ is the output of the reservoir trained to reproduce the $i$-th past input and $cov^2()$ is the covariance between two vectors. The linear short-term memory capacity is then defined as:
\begin{equation}
    MC_{lin} = \sum_{i=1}^\infty m_i    
\end{equation}
The input stream $y_{exp}$ in our case is a random stream of bits. Similar to the $MC_{lin}$ measure, we defined three nonlinear memory capacities. The formulas remain exactly the same, but the training objective changes. These three nonlinear memory capacities are:
\begin{itemize}
    \item $XOR(aa)$, where the reservoir is trained on the XOR of two consecutive bits $aa$.
    \item $XOR(aba)$, where the reservoir is trained on the XOR of two bits $a$ separated by one bit $b$.
    \item $XOR(abba)$, where the reservoir is trained on the XOR of two bits $a$ separated by two bits $b$.
\end{itemize}

\begin{figure}[!t]
\centering
\includegraphics[width=.75\columnwidth]{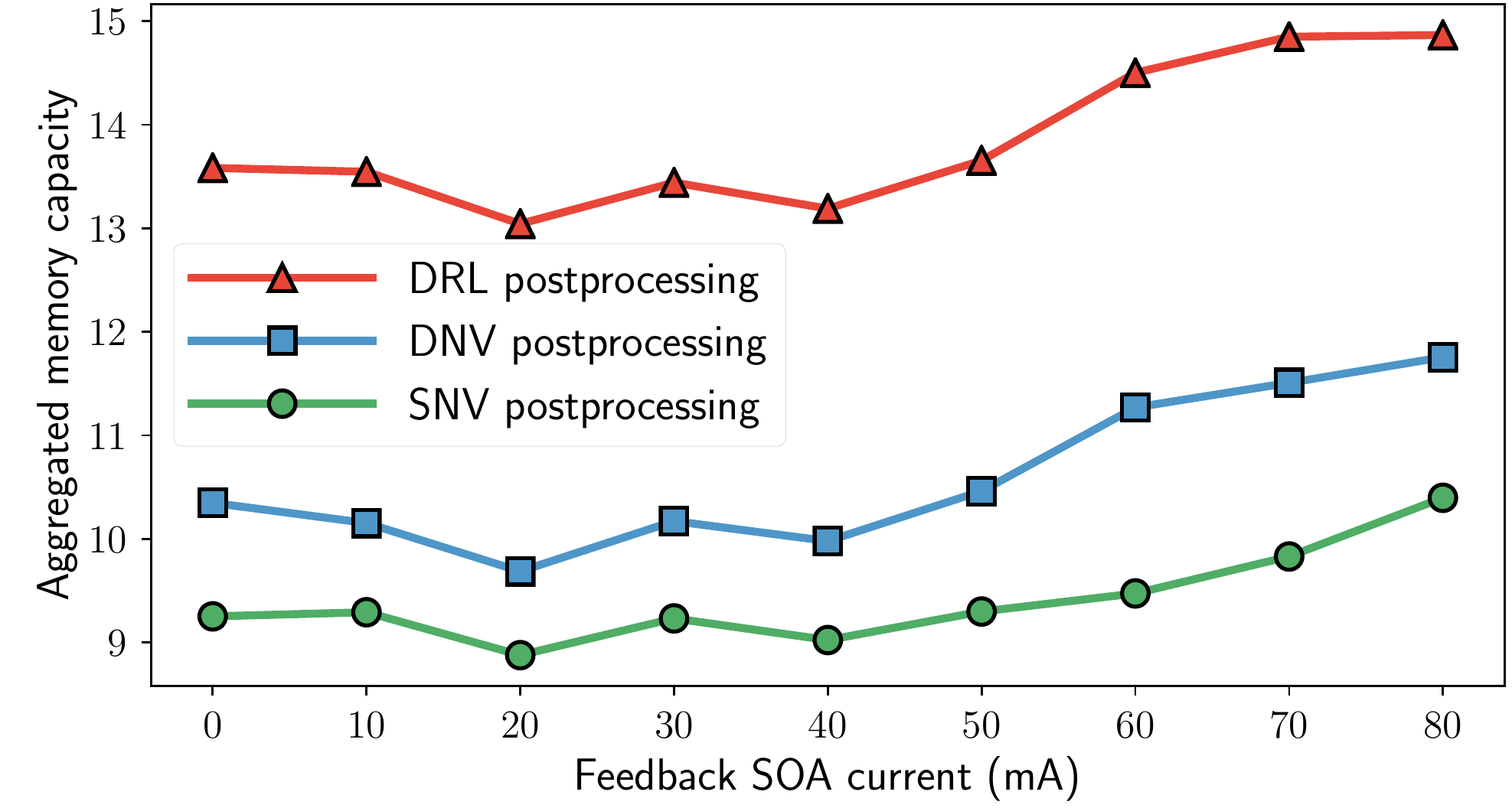}
\caption{This plots illustrates the sum of all capacities shown in Fig.~\ref{fig:MemCap_Var}. We see a clear increase in memory capacity as the feedback increases. The DRL post-processing scheme has the best capacity over the whole range, followed by the DNV post-processing scheme and lastly the conventional SNV post-processing.}\label{fig:MemCap_Total}
\end{figure}

The results for the different memory capacities are shown in Fig.~\ref{fig:MemCap_Var} for varying feedback strengths along the x-axis. For the linear memory capacity $MC_{lin}$, the SNV and DNV post-processing have a value of about 6. This again strengthens the point that there is no need for additional amplification in the feedback line. The DRL post-processing has a linear memory capacity of 8, which is one higher than the expected 7, when comparing to SNV and DRL. These values are around the same value found by Bueno \textit{et al.} \cite{bueno2017conditions} and considerably higher than the linear memory capacity of $2$ found by Takano \textit{et al.} \cite{takano2018compact}. However, the effect of feedback on the linear memory capacity is not very pronounced. For the DRL post-processing the memory capacity $8.25$ and $8.4$ for respectively a feedback SOA current of $0$ and $40$~mA.

For the other memory capacities, we do see a dependence on feedback strength, especially as the distance between first and last bit to be considered, increases. The dependence on feedback is most pronounced for the $MC_{1001}$, as the reservoir has to keep a bit in memory for at least three times the delay time. Hence, we see the link between feedback and the memory inside the system. 

When the individual memory capacities are aggregated, we obtain Fig.~\ref{fig:MemCap_Total}. Here we do observe a dependency on feedback strength. As we mentioned before, feedback strength is in general related to the linear memory capacity, but not exclusively. A task will rarely depend on the linear memory capacity only and in those cases the feedback in the system might still help perform nonlinear transformations over multiple timesteps. It is clear that the DRL post-processing routine has the highest memory capacity, because it has more virtual nodes and it takes the reservoir states, corresponding to the last two masked input data samples, into consideration. The DNV routine outperforms the SNV routine, as it has more virtual neurons per mask-imposed node.

\section{Conclusion}
We have studied the performance of a delay-based reservoir computer, which is designed on a photonic integrated chip. The integrated approach leads to a compact design as well as high computation speeds. We have studied the performance through the Santa-Fe timeseries benchmarking task and we calculated the memory capacity. 

With the conventional reservoir computing scheme, where the mask-imposed nodes coincide with the virtual nodes, we get a performance (best $NMSE=0.135$) which is slightly worse than those found in other works ($NMSE$ around $0.1$). However, we are working in different regimes. While previous works, such as \cite{brunner2013parallel, nguimdo2014fast, bueno2017conditions, takano2018compact}, operate in sub- or near threshold regimes, we operate our laser at pump currents well above the threshold current. We achieve a significant speed up compared to others \cite{paquot2010reservoir, brunner2013parallel}, who achieved speeds in the order of kSa/s and MSa/s respectively. The computation speed of our setup is $0.87$~GSa/s, which is comparable to what Takano \textit{et al.} achieved with additional pre- and post-processing steps.

We were able to improve the performance of the reservoir computer by using different post-processing routines. The first routine is using both readout samples within one mask-imposed node to form the output layer, unlike the conventional routine where we utilize one sample per mask-imposed node. The availability of extra states in the output layer, causes the reservoir computer to perform better. The extra states are not redundant in comparison with the rest, but rather enhance the state space. Since the mask-imposed node has a slightly longer duration than the timescale of the laser, we get two different state values from the transient response on the input. The best performance we achieved here is $NMSE=0.062$.

The second post-processing routines takes the reservoir output for a duration of two delay times. This way we have a richer state space to perform the task and furthermore have access to a longer temporal memory inside this state space, since the last two input data points are present in the two delay times. This post-processing routine has consistently been the best performing out of the three and reaches an $NMSE$ as low as $0.049$. 

We have seen that the best performance for Santa Fe timeseries prediction was found when we the injected signal's wavelength was close to a side-mode, with zero detuning between the injected wavelength and side-mode. We also observed that delay-based RC using semiconductor lasers can achieve very good performances at pump currents well above threshold, where most studies have focused on near-threshold operation. Lastly, we studied the memory capacity of our RC setup as the feedback in the setup is increased and we see a clear increase. Even when the SOAs in the delay line are turned off, we get a linear memory capacity around $8$, which suggests that there is enough feedback already in the system without extra amplification.

\section*{Funding}

Research Foundation Flanders (FWO) (G028618N, G024715N, G029519N); EU Horizon 2020 PHRESCO Grant (688579);  EU Horizon 2020 Fun-COMP Grant (780848); BELSPO IAP P7-35 program Photonics@be; Hercules Foundation; Research Council of the VUB.

\section*{Disclosures}

The authors declare no conflicts of interest.


\bibliography{RC_CHIP.bib} 

\end{document}